\documentclass[sigconf]{acmart}
\newcommand{\name}{MS-RA}
\newcommand{\names}{MS-RA }
\usepackage{array}
\newcolumntype{P}[1]{>{\centering\arraybackslash}p{#1}}

\usepackage{color}

\usepackage[most]{tcolorbox}

\AtBeginDocument{%
  }

\begin{document}

\title{Self-adaptive, Requirements-driven Autoscaling of Microservices}

\author{João Paulo Karol Santos Nunes}
\affiliation{%
  \institution{IBM Brazil and University of São Paulo}
  \country{Barzil}
}
\email{joao.paulo.ksn@gmail.com}

\author{Shiva Nejati}
\affiliation{%
  \institution{University of Ottawa}
  \country{Canada}
}
\email{snejati@uottawa.ca}

\author{Mehrdad Sabetzadeh}
\affiliation{%
    \institution{University of Ottawa}
  \country{Canada}
}
\email{m.sabetzadeh@uottawa.ca}

\author{Elisa Yumi Nakagawa}
\affiliation{%
  \institution{University of São Paulo}
  \country{Brazil}
}
\email{elisa@icmc.usp.br}
\begin{abstract}
Microservices architecture offers various benefits, including granularity, flexibility, and scalability. A crucial feature of this architecture is the ability to autoscale microservices, i.e., adjust the number of replicas and/or manage resources. Several autoscaling solutions already exist. Nonetheless, when employed for diverse microservices compositions, current solutions may exhibit suboptimal resource allocations, either exceeding the actual requirements or falling short. This can in turn lead to unbalanced environments, downtime, and undesirable infrastructure costs. We propose \name, a self-adaptive, requirements-driven solution for microservices autoscaling. \names utilizes service-level objectives (SLOs) for real-time decision making. Our solution, which is customizable to specific needs and costs, facilitates a more efficient allocation of resources by precisely using the right amount to meet the defined requirements. We have developed \names based on the MAPE-K self-adaptive loop, and have evaluated it  using an open-source microservice-based application. Our results indicate that \names considerably outperforms the horizontal pod autoscaler (HPA), the industry-standard Kubernetes autoscaling mechanism. It achieves this by using fewer resources while still ensuring the satisfaction of the SLOs of interest. Specifically, \names meets the SLO requirements of our case-study system, requiring at least 50\% less CPU time, 87\% less memory, and 90\% fewer replicas compared to the HPA. 
\end{abstract}



\maketitle

\section{Introduction}
Microservices have gained considerable popularity in recent years due to their ability to decompose complex applications into multiple  components, i.e., ``small'' applications with a single responsibility that can be tested, deployed, and scaled independently~\citep{7030212}. Microservices further allow flexible management, including individualized scaling operations. Microservices are a close fit for cloud-based applications primarily due to containerization~\citep{Containerized}. A major challenge for developing microservice-based applications is the selection of efficient  autoscaling approaches~\citep{9070951,DBLP:journals/tsc/ChenB17}. Autoscaling refers to the dynamic adjustment of the number of replicas and/or resource management, which can be implemented either \emph{vertically} or \emph{horizontally}. Vertical autoscaling refers to the capacity to automatically increase or decrease resources that applications use, thereby adapting resource usage to the application's needs~\citep{autoscaling}. Horizontal autoscaling  refers to replicas of the same service being added or removed on demand~\citep{Lopez:01}. Scaling microservice-based applications is a complex task. Notably, autoscaling solutions, in addition to allocating resources, need to ensure that the resources are being utilized  effectively. For instance, scaling hardware resources for a bottleneck server without also adjusting associated software resources (e.g., server threads and database connections) can lead to undesirable  response-time fluctuations~\citep{9714008}. If autoscaling is not managed carefully, microservice-based applications may experience resource wastage, an unbalanced environment, downtime, and elevated operational costs.

A recent study on  microservices autoscaling~\citep{Nunes21SAMA} identifies a wide range of approaches with different characteristics and elements, including reactive and predictive operations, orchestration frameworks (e.g., Kubernetes and Docker Swarm), cluster deployment, and machine learning (ML) techniques. Most research in microservices autoscaling introduces new algorithms and enhancements to established methods~\cite{S03,S06,S10}, often tailored to specific microservice types like IoT or web applications~\cite{S04,S07,S08} or focused on managing specific resources such as CPU or memory~\cite{S01,S13,S17,NejatiASB12}. However, existing approaches do not provide a comprehensive solution that simultaneously supports  multiple requirements arising from the application domain, server technologies, clients, and service-level agreements.

In this paper, we propose MicroService, Requirements-driven Autoscaling (\name), a  self-adaptive approach for microservices autoscaling. In \name, the primary focus is on \emph{performance requirements}, which are expressed using service-level objectives (SLOs). Each SLO is a measurable criterion, specifying the targeted level of quality for a specific dimension of performance such as failure rate or rate of responded requests. \names builds upon the well-known MAPE-K self-adaptation loop~\cite{kephart:03,Moreno:15} to periodically monitor metrics necessary to assess user-defined SLOs. The monitoring process checks whether each SLO has been (i) met, (ii) exceeded, or (iii) fallen short of the desired targets. These three levels allow for a ``Goldilocks'' approach, ensuring the efficient allocation of resources, using just the right amount -- neither more nor less -- to meet the SLOs. When it is determined that the system falls short or exceeds certain SLOs, \names scales up or down resources in the case of vertical scaling, and increases or decreases replicas in the case of horizontal scaling, respectively.

We have implemented \names and applied it to an open-source benchmark cloud-based system: the Sock Shop application~\cite{SockShop}. We empirically compare \names with the horizontal pod autoscaler (HPA) -- the industry-standard autoscaling mechanism provided by Kubernetes~\cite{HPA}. Our results show that \names outperforms HPA, using considerably less resources while maintaining the satisfaction of SLOs. Specifically, when the Sock Shop benchmark is subjected to high and fluctuating user-request loads, \names requires at least 50\% less CPU time, 87\% less memory, and 90\% fewer replicas compared to HPA. We provide our complete replication package online~\citep{Anonymized2023}.

\noindent\textbf{Novelty.} The novelty of \names is in shifting the focus of autoscaling from pursuing optimal performance to aligning with specific performance requirements. Unlike traditional methods, our approach not only addresses performance shortcomings but also penalizes excessive resource usage. Grounded in the recognition that business success is often tied to service-level agreements and the effective management of expectations and responsibilities, our autoscaling strategy promotes the \emph{precise} fulfillment of performance requirements. This ensures a more nuanced adaptation that remains mindful of cost trade-offs.

\noindent\textbf{Significance.} 
\names is the result of a collaboration with a leading multinational technology company (name redacted for double-anonymous review) with the goal of addressing autoscaling challenges for the company's diverse clientele, including various cloud providers. While the clients have diverse autoscaling needs, an increasingly common theme is the demand for autoscaling resource-intensive AI technologies, such as large language models. Recognizing the potential cost barriers to optimal performance, clients are now actively seeking a balance between efficiency and expenses. \names provides the first operationalization and initial evaluation of an autoscaling framework designed to help achieve such a balance based on SLO requirements.


\section{The \names Approach}\label{sec:architecture}

\begin{figure}[t]
    \centering
    \includegraphics[width=.45\textwidth]{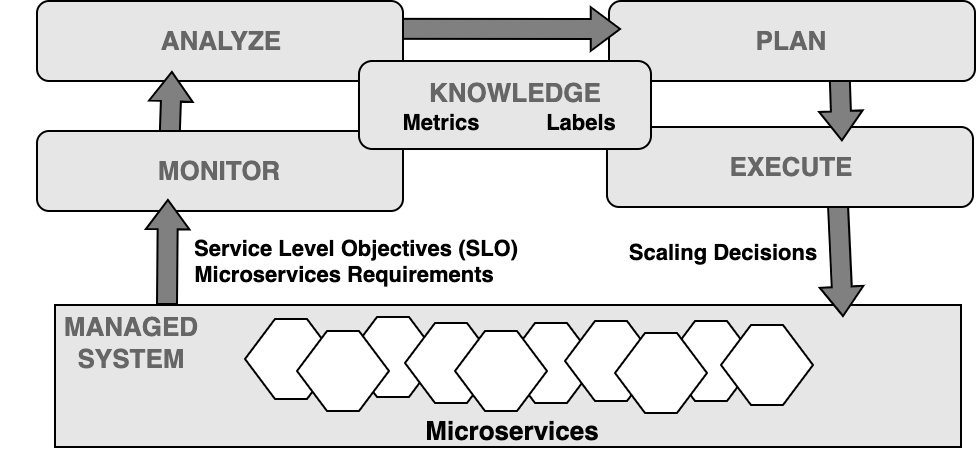}
    \caption{Overview of \name}
    \vspace*{-1em}
    \label{fig:overview}
    \vspace*{-1em}
\end{figure}

Figure \ref{fig:overview} presents an overall view of \name, leveraging the MAPE-K self-adaptation loop, where the managed system consists of a set of microservices. In the following, we present \names by discussing its inputs and  the adapted MAPE-K steps.

\noindent\textbf{Inputs:} \names requires two inputs: (1) SLOs and (2) microservices requirements, 
as described below: 

\noindent\emph{(1) SLOs} are specific, measurable targets related to the performance and reliability of a service. They are often part of a service-level agreement (SLA), which is a formal contract between a service provider and a customer. The SLA outlines the expected level of service, and the SLOs provide detailed, quantifiable targets that help to measure whether the service is meeting those expectations. To illustrate, the following constitutes an SLO, which we will denote hereafter as $\varphi$: \emph{``85\% of requests shall be responded to within 2 seconds over a time frame of 10 minutes.''}

As inputs to \name, customers provide a set of SLOs along with a preferred strategy to achieve each SLO. Strategies are specified in three levels: \emph{conservative}, \emph{normal}, and \emph{best effort}. Below, we define each of these strategy levels and illustrate each, in the context of the example SLO, $\varphi$,  above:

\begin{list}{\labelitemi}{\leftmargin=1em}

\item \textbf{Conservative strategy:}  For this strategy, the autoscaling objective is to keep the SLO metric at 10\% above the threshold specified in the SLO. In our example, the SLO $\varphi$  sets the threshold at 85\%. Hence, a conservative strategy would aim for 95\%. While this strategy tends to utilize more resources, it increases the likelihood of consistently meeting or exceeding the SLO.

\item \textbf{Normal strategy:} This strategy seeks a compromise by keeping a minimum room of 5\% over the threshold. For $\varphi$, this strategy would aim for $90$\%. 

\item \textbf{Best-effort strategy:} This strategy conserves resources by scaling only when the system is below its allowed error-budget limit, which is the same as the established SLO threshold without any margin. For $\varphi$, this strategy would target an 85\% rate for responded requests.  While violations of the SLO would sometimes be inevitable, this strategy is specifically designed to minimize the resources required for operating microservices.

\end{list}

\names regulates the autoscaling process according to the error budget. For our example,  $\varphi$, the error budget is the difference between the current rate of requests responded to by the system and the response rate threshold promised by $\varphi$, i.e., $85$\%. 
When \names fails to meet some SLO, it forces the conservative strategy. Otherwise, it selects one of the three strategies mentioned above based on the remaining margin until \names falls short of meeting some SLO. If the margin is wide, it uses either normal or best effort, depending on customer-specified preferences. If the margin is tight, it resorts to the conservative strategy to minimize the chance of SLO violations. 



\vspace{.1cm}
\noindent\emph{(2) Microservices requirements} identify both the types of servers and the configuration details necessary to facilitate autoscaling. Specifically,  microservices requirements include the following elements: 

\begin{list}{\labelitemi}{\leftmargin=1em}

\item \textbf{Server type:} It defines the type of server on which the microservices are deployed. The server type has a wide range, such as,  web servers, database servers, and application servers.

\item \textbf{Optimizing startup time:} The server type influences resource consumption during container startup, with expensive startups potentially impacting autoscaling efficiency. Some servers initially require substantial resources, which decrease significantly post-startup, resulting in excess unused resources during execution. Therefore, understanding server types and configurations is crucial to optimize the trade-off between startup time and required resources, ensuring effective autoscaling.

\item \textbf{Scaling requirements:} 
Stateless microservices are, in essence, ready to be scaled both horizontally and vertically. However, in real-world scenarios, various contextual factors arise, and it is important to examine whether the server will perform better with horizontal or vertical scaling. As for stateful microservices, it is often the case that these microservices scale better vertically than horizontally due to the need to maintain state between user requests. Yet, there is no one-size-fits-all rule, necessitating an explicit specification of scaling requirements.

\item \textbf{Custom metrics:} 
In addition to the standard metrics, such as response time, throughput, and latency, \names allows one to define custom metrics  to be collected for monitoring. These include connection time and message-response time.
\end{list}


\noindent\textbf{Monitor:} 
This step is responsible for systematically monitoring the microservices logs and extracting SLO metrics, such as response time, downtime, and custom user-defined metrics. Specifically, this step performs the following two tasks: (i)~data collection, which records the metric values pertinent to SLOs in a time-series format, and (ii)~data cleaning, which ensures that the data adheres to a consistent schema and is duplicate-free.

\noindent\textbf{Analyze:}  This step classifies each vector of metrics collected in the Monitor step into three categories: \emph{met}, \emph{exceeded}, or \emph{poor}. This classification is a holistic verdict on the satisfaction of all SLOs. 
Specifically, \emph{poor} indicates that the system violates some SLO; \emph{met} indicates that the system meets all SLOs; and, \emph{exceeded} indicates that the system meets all SLOs but exceeds some. This means that the system has overshot the target with respect to some SLO and can potentially be wasting cluster resources.


\noindent\textbf{Plan:} This step identifies an autoscaling strategy when the system status is labelled either as \emph{poor} or \emph{exceeded} in the Analyze step.  The purpose of this step is to select among the three strategies, discussed earlier (i.e., conservative, normal, and best effort) and choose one that can effectively guide the system towards achieving the \emph{met} status. Once the strategy is selected, \names uses reactive, threshold-based autoscaling~\cite{Nunes21SAMA} for adaptation. Unlike previous threshold-based autoscaling approaches, \names defines thresholds based on SLOs and customer metrics. Further, the transition between alternative strategies is adaptively managed, guided by a dynamically computed  error budgets.

\noindent\textbf{Execute:} This step applies the autoscaling strategy identified in the Plan step to the managed system. To execute the autoscaling strategy, this step directly uses the cluster API server, asking for changes in the microservices scaling. The changes are applied using the rolling updates strategy~\cite{RollingUpdate}, which allows updates to deployments to occur with zero downtime by incrementally replacing pod instances with new ones.

\section{Empirical Evaluation}\label{sec:evaluation}

We evaluate \names through the following research question:

\textbf{RQ}: \emph{How effectively can \names meet SLOs by allocating resources to microservices using just the right amount and without exceeding needs?}
To answer RQ, we apply \names to the Sock Shop application -- an open-source system widely used in the literature as a benchmark for microservice-based systems. To demonstrate the effectiveness of \name, we compare it with HPA (horizontal pod autoscaler), which is a standard industry-strength autoscaling mechanism in Kubernetes.

\subsection{Implementation of \names}
Our base environment is the Openshift containerization platform \cite{Openshift}, which is built on top of Kubernetes. \names is implemented in Python using Ansible operators. For data collection, we use Prometheus~\cite{Prometheus}, which collects and stores metrics as time-series data. This means that each piece of metric information is timestamped to indicate when it was recorded. The data is stored after classification into MongoDB~\cite{MongoDB}. Load testing is implemented using Apache JMeter -- an open-source tool written in Java, designed to test functional behaviour and measure performance~\cite{JMeter}.

\subsection{Experiment Setup}
\label{subsec:expsetup}

This section describes our study subject, evaluation configurations, and analysis procedure.

\noindent\textbf{Study subject.} The Sock Shop application simulates the user-facing part of an e-commerce website that sells socks. The application features a front-end microservice that acts as a proxy for all HTTP client requests, rerouting these calls to the appropriate microservices, like user login.  To focus on testing the scalability of the system nodes responsible for business logic, rather than the user interface, we replicated the front-end microservice. This ensures that the business-logic system nodes can receive varying levels of traffic applied to the system.

\noindent\textbf{Configurations.} We configure both \names and HPA to enable the comparison of their performance. For \name, we consider two commonly used SLOs for microservice-based systems: 

\begin{list}{\labelitemi}{\leftmargin=1em}

\item SLO1: \emph{X\% of the calls shall be responded to in 2.5 seconds.} 
\item SLO2: \emph{The failure rate shall be kept below Y\%.}

\end{list}

Recall from Figure~\ref{fig:overview} that \names requires SLOs as inputs. In our experimentation, we supply \names with different values for X in SLO1 and Y in SLO2. The values are detailed in Table~\ref{tbl:msdda1_hpa}. Specifically, we test three sets of values for X and Y, leading to three distinct \names configurations: \name-A, \name-B, and \name-C. For instance, the \name-A configuration aims for 95\% of calls to be answered in less than 2.5 seconds and maintains a failure rate below 0.5\%, whereas the \name-B configuration aims to respond to 90\% of calls within 2.5 seconds with a failure rate under 1\%. The \name-A, \name-B, and \name-C configurations correspond to the conservative, normal, and best-effort strategies, respectively, as discussed in Section~\ref{sec:architecture}.


\begin{table}[t]
\caption{SLO and microservice requirements parameters for \names and autoscaling input parameters for HPA (baseline)}~\label{tbl:msdda1_hpa}
\small
\scalebox{0.95}{\begin{tabular}{|p{1.4cm}|P{0.6cm}|P{0.6cm}|P{2cm}||P{1cm}|P{0.7cm}|}
\hline
\textbf{\names Configs} & \textbf{SLO1 (X)} & \textbf{SLO2 (Y)} & \textbf{Vertical scaling CPU and Memory rates} & \textbf{HPA Configs} & \textbf{CPU thre\-shold} \\ \hline\hline
\name-A & 95\%           & 0.5\%            & 20\%   & HPA-A   & 60\%          \\
\name-B & 90\%           & 1\%              & 10\%   & HPA-B   & 70\%          \\
\name-C & 85\%           & 2\%              & $0\%$  & HPA-C   & 80\%           \\ \hline
\end{tabular}}
\end{table}

Recall from Figure~\ref{fig:overview} that \names, in addition to SLOs, also requires microservices requirements as part of its inputs. Specifically, since \names supports both vertical and horizontal scaling, it requires vertical CPU and memory usage rates as input. These rates determine the amount of resources allocated or removed for each replica during scaling up or down. Table~\ref{tbl:msdda1_hpa} shows the vertical scaling CPU and memory rates used for the \name-A, \name-B, and \name-C configurations in our experiments.

In contrast to \name, the HPA autoscaling baseline requires thresholds for CPU usage as input triggers for autoscaling. HPA performs scaling when the CPU usage exceeds the specified thresholds. Similar to the inputs for \name, we consider three alternative threshold values for HPA, as shown in Table~\ref{tbl:msdda1_hpa}. We refer to the configurations of HPA based on these threshold values as HPA-A, HPA-B, and HPA-C. Since HPA does not support vertical scaling, the vertical scaling rates do not need to be indicated for HPA.

\noindent\textbf{Analysis procedure.}
We create streams of input requests  to mimic actual usage of the Sock Shop application. We assume input requests are sent over a duration of 30 minutes with varying numbers of users. Specifically, we divide the 30-minute duration into time intervals of five minutes. During the first five minutes, we have 10 users sending requests to the application. During the second, third, fourth, fifth, and sixth intervals, we create 20, 30, 10, 30, and 20 users, respectively, sending requests. We apply the above streams of user requests to the Sock Shop application for each configuration of \names and HPA in Table~\ref{tbl:msdda1_hpa}. We repeat each experiment ten times to account for random variation. 

\subsection{Results}
In our comparison, we aim to determine which autoscaling technique, \names or HPA,  can always ensure SLO1 and SLO2 while minimizing resource  utilization. Resource utilization includes the CPU and memory consumption and the number of pod replicas. 

\begin{figure*}[!t]
    \centering
    \includegraphics[width=14.5cm]{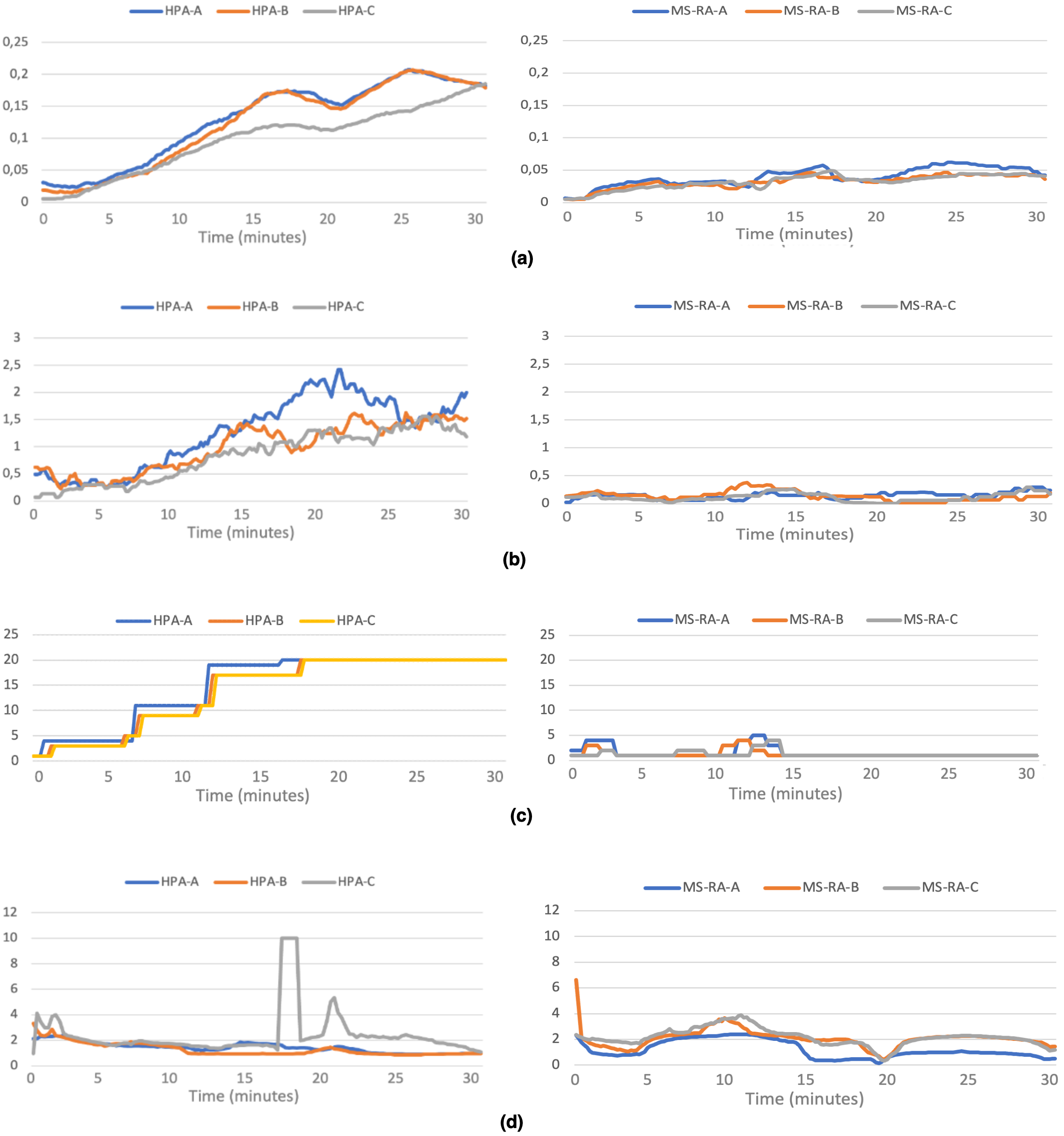}
    \caption{Comparing the configurations of \names and HPA described in Table~\ref{tbl:msdda1_hpa}: (a) CPU consumption comparison (Millicpu); (b) Memory consumption comparison (Megabytes); (c) Number of replicas comparison; and (d) Response time comparison.
    }
    \label{fig:results_graphics}
\end{figure*}

Figures \ref{fig:results_graphics}(a)-(c) respectively show the CPU consumption, memory consumption  and the number of replicas obtained by  applying a varying number of user requests for 30 minutes, as described in our analysis procedure, to three different configurations of \names and HPA. As shown in Figures \ref{fig:results_graphics}(a)-(b), \names uses considerably less CPU and memory compared to HPA.  This shows that the SLO-driven autoscaling of \names can substantially reduce CPU and memory usage compared to HPA. 


As for the number of replicas, Figure \ref{fig:results_graphics}(c) shows that HPA keeps increasing the number of replicas and fails to scale down even when the input load is reduced.  We note that HPA requires as input a stabilization-time parameter, which is a delay duration between a scale-down trigger and the execution of the scaling-down action. For our experiments, the stabilization-time parameter is set to less than two minutes for HPA-A and HPA-B and to less than one minute for HPA-C. However, Figure \ref{fig:results_graphics}(c) shows that the number of replicas has remained at 20 for a period much longer than the stabilization time, without any indication of going down and thus providing no evidence of replica reduction, even if we were to choose a longer experiment duration. In contrast, \names is able to effectively both scale up and down the number of replicas, ensuring that the quantity and active duration of these replicas are maintained at the required level to satisfy the underlying SLOs. 

An important observation from Figure~\ref{fig:results_graphics}(c) is that the overall number of replicas that \names generates is much lower than that generated by HPA. Specifically, for \name, the replica numbers stabilize at one. This is because \names combines adaptive vertical scaling with horizontal scaling. In this case, adaptive vertical scaling of \names increases CPU capacity with each new replica, resulting in the final remaining replica having greater capacity than initially planned, enabling it to handle the whole load. 

Figure \ref{fig:results_graphics}(d)  compares response times for HPA and \name. Overall, HPA has a lower response time than \names. However, while \names might not offer the lowest response time, it ensures that response time remains at a level such that SLO1 is met most of the time. This is consistent with the adaptive SLO-driven strategy of \name, as it attempts to limit resource consumption while minimizing or eliminating  SLO violations.

\begin{tcolorbox}[colback=gray!10!white,colframe=black!75!black]
To answer \textbf{RQ}, Table \ref{tab:results} presents the average number of replicas, CPU, and memory usage for all HPA and \names configurations as defined in Table~\ref{tbl:msdda1_hpa}. In addition, Table~\ref{tab:results} includes the number of SLO1 and SLO2 violations observed throughout the duration of the experiments. Overall, while HPA and \names show comparable performance in terms of SLO violations, \names configurations achieve substantially lower resource utilization compared to those of HPA. Notably, the most effective \names configuration, \mbox{\name-A}, manages to maintain zero SLO violations while requiring at least 50\% less CPU time, 87\% less memory, and 90\% fewer replicas compared to HPA.
\end{tcolorbox}

\begin{table}[t]
\centering
\caption{Summary of the results comparing different configurations of \names and HPA described in Table~\ref{tbl:msdda1_hpa}} \label{tab:results}
\small
\scalebox{0.95}{\begin{tabular}{|l|P{1.1cm}|P{1cm}|P{1cm}|P{1cm}|P{0.8cm}|}
\hline
\textbf{Profile} 
& \textbf{Average number of replicas}      
&\textbf{CPU (millicpu)}
&\textbf{Memory (Mega\-bytes)}
&\textbf{SLO1 Violations}
&\textbf{SLO2 Violations}
\\ \hline\hline
HPA-A & 14.94 & 0.128 & 1.254 & 0 & 0 \\
HPA-B & 13.86 & 0.123 & 0.984 & 1 & 0 \\
HPA-C & 11.51 & 0.097 & 0.833 & 2 & 0 \\
\textbf{\name-A} & \textbf{1.51} & \textbf{0.039} & \textbf{0.145} & \textbf{0} & \textbf{0} \\
\name-B & 1.29 & 0.032 & 0.136 & 1 & 0 \\
\name-C & 1.25 & 0.031 & 0.111 & 2 & 0 \\ \hline
\end{tabular}}
\end{table}

\subsection{Threats to Validity}
\textbf{Internal validity.} To enhance the internal validity of our study, we implemented measures to ensure that our metric readings remained unaffected by factors unrelated to our experiments. Specifically, we ensured that the number of concurrent users created in our experiments was selected in a way that would avoid causing network congestion. We also continuously monitored the health of the network throughout our experiments to verify that no congestion arose due to circumstances beyond our control. Furthermore, we repeated our experiments ten times and at different times of the day  to ensure that our results are representative and not impacted by other processes running  on the cluster.

\textbf{External validity.} While our experimental setup builds upon standard and widely used practices, we note that our evaluation was performed on a single benchmark system. To more conclusively examine the generalizability of our findings and improve external validity, we intend to conduct further experiments using various cloud-based applications and different microservices technologies. 

\begin{table}[]
\centering
\small
\caption{Comparison of \names and related work} \label{tab:relatedwork}
\scalebox{0.9}{\begin{tabular}
{|p{4.2cm}|
p{0.16cm}|p{0.16cm}|p{0.16cm}|p{0.16cm}|p{0.16cm}|p{0.16cm}|p{0.16cm}|}
    \hline
        \textbf{Related Work} & 
        \rotatebox[origin=c]{90}{\textbf{Custom metrics}} & \rotatebox[origin=c]{90}{\textbf{SLO awareness}} & \rotatebox[]{90}{\textbf{ Server-type awareness }} & \rotatebox[origin=c]{90}{\textbf{Self-adaptivity}} & \rotatebox[origin=c]{90}{\textbf{Quality assurance}} & \rotatebox[origin=c]{90}{\textbf{Extensibility}} & \rotatebox[origin=c]{90}{\textbf{Domain-agnostic}} \\ \hline\hline
        \citet{9714008} &  
        & \checkmark & \checkmark &  &  &  &  \\ \hline
        \citet{electronics12010240} & 
        &  & &  & \checkmark &  &  \\ \hline
        \citet{10.1145/3502181.3531460} & 
        & \checkmark &  &  & \checkmark &  & \checkmark \\ \hline
        \citet{9932943} & 
        &  &  &  &  &  & \checkmark \\ \hline
        \citet{9036958} & 
        &  &  &  &  &  &  \\ \hline
        \citet{8885153} &  
        & \checkmark &  & \checkmark &  &  &  \\ \hline
        \citet{Pozdniakova} & 
        \checkmark &  &  & \checkmark &  &  & \checkmark \\ \hline
        \name & 
        \checkmark & \checkmark & \checkmark & \checkmark & \checkmark & \checkmark & \checkmark \\ \hline
    \end{tabular}}
    \vspace*{-1em}
\end{table}
\section{Related Work}\label{sec:discussion}

Table~\ref{tab:relatedwork} provides a comparative analysis between our solution, \name, and recent related work. Our comparison emphasizes the key  characteristics of our autoscaling solution, namely, the ability to support custom metrics, guidance by SLOs, awareness of various server types, self-adaptivity, provisions for quality assurance, and the solution's extensibility and domain independence.

\citet{Pozdniakova} takes \textbf{custom metrics} into account but without covering how these metrics should be part of the solution or offering an evaluation.
There are approaches (e.g., \citep{9714008,10.1145/3502181.3531460,8885153}) that are \textbf{SLO-aware} in the sense that they monitor SLOs. However, these approaches do not make adaptation decisions based on SLOs. Unlike these earlier approaches, \name\ is fundamentally \emph{SLO-driven}, with its goal being to precisely meet the given SLOs without exceeding or falling short. Furthermore, the  earlier approaches do not consider different strategies such as conservative, normal and best effort in order to optimize the use of resources.

\citet{9714008} develop a \textbf{server-type-aware} approach to support the adjustment of hardware and software resources (e.g., server threads or connections) at runtime. However, in contrast to our work, this earlier study does not address dynamic microservices environments such as Kubernetes or the optimization of the resources required to initialize the servers.

Similar to \name, \citet{8885153} and \citet{Pozdniakova} provide \textbf{self-adaptive} autoscaling. However, in contrast to \name, they do not support vertical scaling capabilities, making them unsuitable for handling stateful microservices. Our approach is an adaptive solution that covers proactive/reactive and vertical/horizontal scaling strategies. It also supports stateful microservices that often need to scale vertically. By default, our approach can identify a suitable strategy, but users can also specify their preferred strategy upfront, to be used when there are multiple sensible options at a given time. Furthermore, users are given the flexibility to disable options that are not of interest. For example, in the case of stateful microservices, horizontal scaling can be disabled.

As for supporting \textbf{quality assurance}, \citet{electronics12010240} propose an infrastructure for workload monitoring and ensuring quality-of-service (QoS) requirements for the entire cluster hosting a system. However, this infrastructure lacks  fine-grained mechanisms for handling individual microservices. In a similar vein, \citet{10.1145/3502181.3531460} employ a conservative QoS strategy, initially allocating ample resources to meet SLOs for all microservices, and then iteratively attempting  resource reduction based on application-performance statistics. \names provides monitoring at the level of \emph{individual} microservices, ensuring the efficiency of the chosen autoscaling method. In addition, \names can employ not only normal and conservative strategies but also a best-effort strategy.

Regarding \textbf{extensibility}, to our knowledge, \names stands out due to its support for being instantiable with different autoscaling algorithms. This aspect is crucial for maintaining the relevance of our solution, given that the field of microservices autoscaling is continuously evolving.

We know of three earlier studies~\cite{10.1145/3502181.3531460,9932943,Pozdniakova} that share with \names the characteristic of being \textbf{domain-agnostic}. The main advantage of \names over these earlier studies is that it offers more flexibility for domain experts to customize server-type configurations. The unique combination of being domain-agnostic and server-type aware enables \names to be tailored with specific domain parameters, maintaining its versatility across various domains.

In summary, and as shown by Table~\ref{tab:relatedwork}, when compared to the state of the art, \names offers a more comprehensive solution for scaling microservices across multiple important dimensions. Furthermore, as demonstrated by the evaluation in Section~\ref{sec:evaluation}, our solution outperforms HPA, representing the current state of practice.


\section{Conclusions and Future Research}\label{sec:finalconsiderations}
Microservices architecture has drawn considerable attention from the research community in recent years and has further gained popularity in many companies, including prominent ones like Amazon, Netflix, Uber, and eBay. Many challenges nonetheless remain in implementing microservices architecture, including the difficulty of microservices management and the lack of an efficient microservices autoscaling solution. Our work positions itself in this context by introducing \name, an SLO-driven self-adaptive autoscaling solution. The primary goal of \names is to ensure the efficient allocation of resources, using just enough to meet the service-level objectives (SLOs) of interest without any waste.

We elaborated on how \names adapts the standard MAPE-K self-adaption loop by detailing the inputs required by \names and its specific steps. Furthermore, we described how \names is implemented and applied to a benchmark system. We provided an evaluation of \names by comparing it with a popular industrial baseline, the horizontal pod autoscaler (HPA). Our evaluation results indicate that \names fulfills its goal by curbing resource wastage while simultaneously keeping performance and failure rates within the desired margins. Specifically, the results show that, for our benchmark system, a suitable configuration of \names achieves zero SLO violations while requiring at least 50\% less CPU time, 87\% less memory, and 90\% fewer replicas compared to HPA.

The research reported here represents the initial step in a longer-term agenda, aiming to develop requirements-driven autoscaling and improve resource utilization efficiency in microservices architecture. In future work, we aim to broaden our evaluation to encompass larger applications and explore systems from diverse domains. 
In addition, we envision enhancing the existing threshold-based autoscaling algorithm of \names with a proactive component powered by a suite of machine learning techniques. Through the fusion of machine learning algorithms and threshold-based algorithms, our ultimate goal is to enhance \names so that it can transition seamlessly between reactive and proactive modes according to demand, while ensuring the satisfaction of SLOs.


\begin{acks}
We gratefully acknowledge funding from FAPESP (2023/00488-5), CNPq (313245/2021-5), 
and NSERC of Canada under the Discovery and Discovery Accelerator programs. 
\end{acks}

\bibliographystyle{ACM-Reference-Format}
\balance
\bibliography{references}

\end{document}